# Record Responsivity–Conductance Performance in Sub-bandgap-triggered Ga$_2$O$_3$ PCSS

Vikash Jangir[1], Sourojit K. Mazumder[2], and Sudip K. Mazumder[1]
[1]Department of Electrical and Computer Engineering, University of Illinois Chicago
[2]Department of Electrical and Computer Engineering, University of Illinois Urbana-Champaign

*Abstract*
This work reports enhanced photoconductive switching performance in Fe-doped β-Ga$_2$O$_3$ PCSS through the concurrent optimization of electrode geometry and optical excitation wavelength. By systematically varying the anode grid pitch (20–80 µm) and excitation spectrum (235–500 nm), we identify a key sub-bandgap regime centered at 272 nm that activates deep-level defect states and enables efficient bulk carrier transport. In contrast to above-bandgap excitation, which is limited by shallow surface absorption, sub-bandgap illumination promotes strong photocurrent generation and improved carrier collection. Under optimized conditions with a 40 µm pitch, the device exhibits a high peak photocurrent of 4.14 A and a record-low on-resistance of 10.4 Ω. To quantify this simultaneous achievement, we define a responsivity–conductance figure of merit ($FoM_{RC}$), which reaches a record value of $4.7 \times 10^{-6}$ S/W. These results underscore the suitability of Fe-doped β-Ga$_2$O$_3$ for next-generation high-power optoelectronic switching, enabling robust ampere-level photocurrents together with low on-resistance through optimized device geometry and sub-bandgap excitation.

## 1. Introduction

Photoconductive semiconductor switches (PCSSs) yield ultrafast switching dynamics by optically modulating the conductivity of a bulk semiconductor substrate rapidly [1][2]. Furthermore, the optical excitation of a PCSS prevents backpropagation from a high-voltage power stage to a low-voltage microelectronics control unit, which is a common issue in typical switching power-electronics systems due to high voltage and current slew rates [3]. As such, PCSSs have wide applications, including, but not limited to, pulsed power, fly-by-light, solid-state circuit breakers, kinetic weapons, electromagnetic launch systems, high-power radio-frequency power amplifiers, etc. The confluence of ultra-wide-bandgap (UWBG) semiconductors [4], which have higher critical electric-field strengths that yield high-voltage operation and lower device dimensions (and capacitances), and PCSS is expected to provide significant impetus to the growth and utility of these next-generation ultrafast optically-triggered power semiconductor technologies.

Among the UWBG semiconductors, gallium oxide (Ga$_2$O$_3$) [5] is considered one of the most promising for near-term commercialization, primarily due to its potential for large-area substrates, low cost, low threading dislocation density (TDD), and capability for large-scale production. High-quality Ga$_2$O$_3$ epilayers are now possible using MOCVD, HVPE, and MBE, with excellent wide n-type doping control and low background compensation. As such, high-voltage electrical power semiconductor devices like power MOSFETs and diodes are already being pursued, and preliminary demonstrations have been conducted [6], [7]. Recently, experimental research on Ga$_2$O$_3$ PCSS, utilizing a longer wavelength optical beam and low current and voltage, has been initiated [8], which illustrates the preliminary potential of this UWBG material and highlights the need for further exploration in this promising area.

This work investigates the impact of varying the anode electrode grid pitch separation (which also serves as the optical window) and the wavelength of optical excitation on maximizing the responsivity-conductance trade-off in a Fe-doped β-Ga$_2$O$_3$ PCSS. The experimental results demonstrate that Fe-doped β-Ga$_2$O$_3$ PCSS supports ampere-level current conduction and record-low on-resistance under medium operating voltages, while extending the photoresponse into longer wavelength regimes. These results highlight the advantages of Fe-doped β-Ga$_2$O$_3$ over previously reported UWBG PCSS devices, demonstrating the former's potential as a cost-effective optical device for next-generation high-power optoelectronic switching applications.

## 2. Device Structure and Fabrication

The β-Ga$_2$O$_3$ PCSS were fabricated using double-side–polished 450 µm thick, Fe-doped semi-insulating β-Ga$_2$O$_3$ (010) substrates obtained from Kyma Technologies with an Fe doping concentration of approximately $1 \times 10^{18}$ cm$^{-3}$. Prior to fabrication, the substrate was sequentially cleaned in acetone, isopropyl alcohol (IPA), and deionized (DI) water, followed by drying with high-purity nitrogen to remove organic residues and surface contaminants. Fig. 1 presents both the schematic cross-section and top-view micrographs of the fabricated devices. The device structure (Fig. 1) comprises a continuous ground electrode on the back surface and a grid-type top electrode on the top surface, allowing direct optical access to the active region. Both electrodes were formed by electron-beam evaporation of a Ti/Au (50 nm/100 nm) bilayer. The top electrode was patterned by image-reversal lithography (AZ 5214E) and lift-off to define a grid geometry consisting of alternating metal fingers and transparent apertures. The devices were subjected to rapid thermal annealing at 480 °C for one minute in a nitrogen ambience to achieve ohmic contact [9].

The grid-like anode electrode of the PCSS produces a predominantly vertical electric field across the illuminated region, while fringing fields near the finger edges provide lateral components that assist in collecting carriers generated close to the surface. To explore the role of electrode geometry, the electrode pitch $g$ [Fig. 1(a)] was systematically varied between 20 and 80 µm. A narrow pitch enhances carrier collection through strong fringing-field overlap and shorter lateral drift paths, whereas a wider pitch increases the optical fill factor but introduces extended low-field regions between fingers that limit transport efficiency. This response exhibits a strong wavelength dependence: short-wavelength UV photons are absorbed within nanometer-scale depths, where transport is dominated by lateral drift and surface recombination. In contrast, sub-bandgap illumination penetrates deeper into the bulk, generating carriers that couple to the vertical electric field. This volumetric excitation makes the photocurrent response highly sensitive to electrode spacing, requiring an optimized pitch to balance optical access with field intensity. Importantly, the pitch variation does not alter the dominant vertical conduction path through the 450 µm substrate but instead modifies the lateral access to vertical field lines, thereby shaping carrier collection dynamics, peak current, and the observed switching speed. Systematic pitch variation, therefore, provides a controlled means of benchmarking device performance limits and deriving design guidelines for robust high-field switching operation.

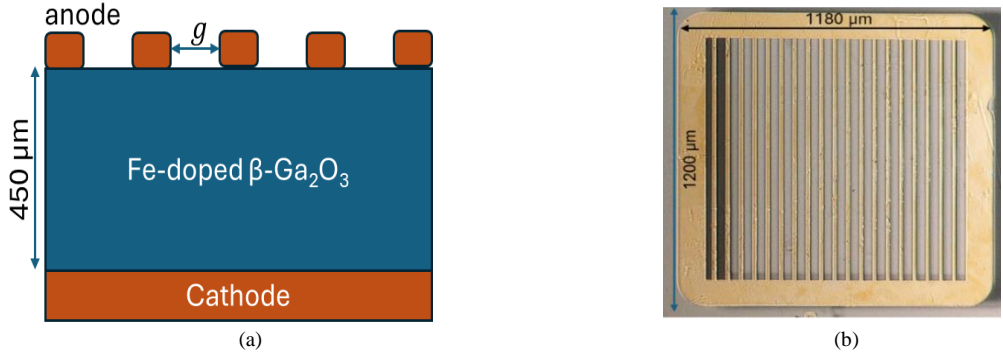

Fig. 1. (a) Schematic cross-section and (b) top view of the fabricated Fe-doped β-Ga₂O₃ PCSS.

### 3. Experimental Test Setup Configuration

A Keithley 2657A source meter, interfaced to a probe station using a Keithley 8020 high-power interface channel, is used to measure the dark current and off-state resistance ($R_{OFF}$). For transient response, the PCSS devices were packaged in an optical open-cavity package using both thermally and electrically conductive epoxy. Connections with the package pins were enabled using wire bonding, as shown in Fig. 2(a). Fig. 2(b) shows the experimental setup used for the optical triggering of the PCSS devices. A 100 nF capacitor is charged to the desired DC bias ($V_{bias}$) through a 1 kΩ current-limiting resistor. All PCSS devices are triggered using an OPOLETTE UX06230U tunable laser with a beam diameter of 3 mm and a full width at half maximum (FWHM) of 7 ns. The intensity of the laser beam varies with respect to wavelength. Once the PCSS device is optically excited, the resulting electrical pulse is measured across a 50-Ω terminating load resistor ($R_{load}$) using a Tektronix TPP1000 voltage probe and a Tektronix MSO46 oscilloscope, both of which have a bandwidth of 1 GHz. The laser pulse profile is simultaneously monitored using a high-speed silicon photodetector (DET10A2, Thorlabs).

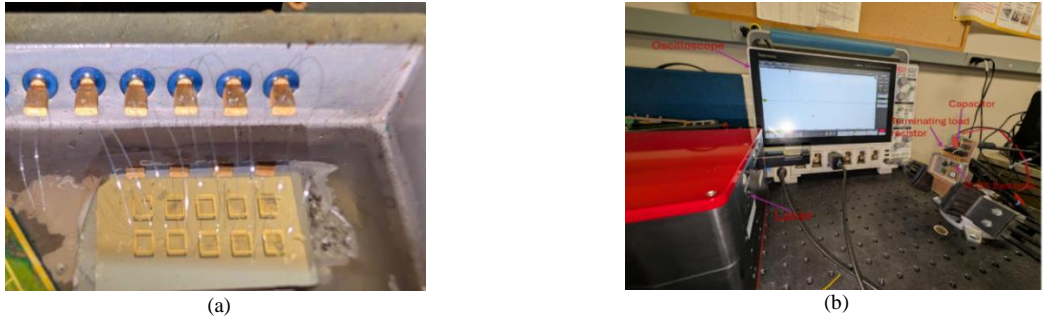

Fig. 2 (a) The packaged Fe-doped β-Ga₂O₃ PCSS. (b) A test set-up for transient characterization of the PCSS.

### 4. PCSS Experimental Results and Analysis

The photo responses of four Fe-doped β-Ga₂O₃ PCSS devices with top electrode pitches of 20, 40, 60, and 80 µm were systematically characterized under a DC bias ($V_{bias}$) of 250 V over the wavelength range of 235–500 nm. These measurements for the PCSS capture both above-bandgap and sub-bandgap response regimes.

The normalized responsivity and the responsivity–conductance figure of merit ($FoM_{RC}$) of Fe-doped β-Ga₂O₃ PCSS devices with different electrode gaps as a function of incident wavelength are shown in Figs. 3(a) and 3(b), respectively. To isolate the intrinsic switching performance from the circuit-limited current saturation, a quantitative metric for switching efficiency, $FoM_{RC}$ is introduced. This metric is defined as the increase in on-state conductance per unit optical power; that is, $FoM_{RC} = G_{ON}/P_{OPT}$, where $G_{ON} \approx 1/R_{ON}$ ($R_{ON}$ is the on-state resistance of the PCSS) represents the net photoconductance and $P_{OPT}$ is the peak incident optical power. All devices displayed a pronounced responsivity peak in the range of 270–275 nm, occurring at wavelengths approximately 12–17 nm longer than the intrinsic absorption edge of β-Ga₂O₃ (~258 nm). This offset suggests that contributions from sub-bandgap processes may play a significant role in this regime, in addition to, or instead of, direct band-to-band transitions. At the intrinsic absorption edge, most incident photons are absorbed within a few nanometers of the PCSS surface due to the large absorption coefficient [10], which limits the effective generation of carriers. The enhanced response near 272 nm can be considered in relation to Urbach tail absorption as well as the possible photoionization of deep-level defect complexes within the semi-insulating substrate [11-12]. Earlier deep-level optical and transient spectroscopy studies have reported several defect states in β-Ga₂O₃, including $E_C$-0.63 eV, $E_C$-0.81 eV, $E_C$-1.29 eV, $E_C$-2.0 eV, and $E_C$-4.48 eV [13–18]. The $E_C$-0.81 eV state is often associated with Fe dopant activation [19, 20]. In addition, Angeloni et al. [21] identified Fe-related states at $E_C$-3.05 ± 0.05 eV and $E_C$-3.85 ± 0.05 eV using picosecond tunable UV spectroscopy. Sub-bandgap excitation around 272 nm may therefore facilitate carrier photoionization from these Fe-related defect states, which could explain the enhanced responsivity observed in this wavelength range.

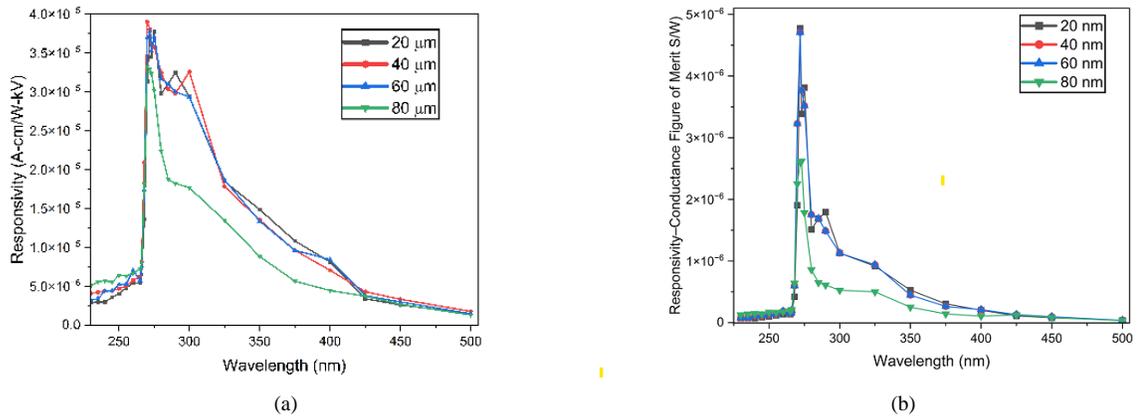

Fig. 3 (a) Normalized responsivity and b) responsivity–conductance figure of merit ($FoM_{RC}$) of the Fe-doped β-Ga₂O₃ PCSS with variation in wavelength.

Fig. 4(a) presents the peak photocurrent ($I_{Peak}$), while Fig. 4(b) shows the corresponding minimum on-resistance ($R_{ON}$) for both above- and sub-bandgap excitations. The photocurrent responses exhibit distinct regimes determined by the carrier generation profiles, arising from the interplay between wavelength-dependent optical penetration depth and the electric-field distribution. For deep sub-bandgap excitation (450 and 500 nm), photocarriers are generated predominantly through volumetric excitation of Fe- and Ir-related deep levels within the bulk. These carriers couple efficiently to the vertical drift field, yielding intermediate photocurrent values (approximately 1–1.7 A at 450 nm and < 1 A at 500 nm). Both wavelengths exhibit a clear maximum at a 40 μm pitch. This trend reflects a trade-off where the active collection volume expands with pitch but is eventually overtaken by the reduction in average electric-field intensity at larger gaps. Under above-bandgap excitation (245–255 nm), carrier generation is confined to a shallow near-surface region. Carrier transport is dominated by lateral diffusion and surface-driven drift, where strong surface recombination suppresses the photocurrent response (< 1 A). The slight increase in photocurrent observed with increasing pitch is ascribed to the geometric expansion of the photo-active surface area, although the dominance of diffusion-limited transport prevents significant scaling.

Near-band-edge excitation (272 and 300 nm) produces strong photocurrent response owing to an optimal penetration depth that simultaneously activates deep levels and maintains strong near-surface generation. Specifically, under 272 nm excitation at a 40 μm pitch, the photocurrent reaches a saturation plateau of ~4.14 A. This value corresponds to the circuit-limited peak current ($I_{peak} = V_{bias}/(R_{load} + R_{ON})$) defined by the 50-Ω load ($R_{load}$) and $V_{bias}$ of 250 V, indicating that the intrinsic $FoM_{RC}$ in this regime is sufficiently high to drive the device into a fully conductive state ($R_{ON}$ ~10.4 Ω), limited only by the external circuit. At 300 nm, the pitch dependence resembles that of the deep sub-bandgap regime, exhibiting a clear peak at 40 μm followed by a sharp drop at 80 μm. For the 272 nm case, the photocurrent remains nearly constant over the 20–60 μm range due to this saturation effect. However, at 80 μm, both near-band-edge wavelengths display a pronounced decline. This reduction is attributed to reduced overlap of lateral fringing fields at the widest anode spacing, which weakens carrier funneling into the high-field vertical region and consequently limits the overall photo-collection efficiency.

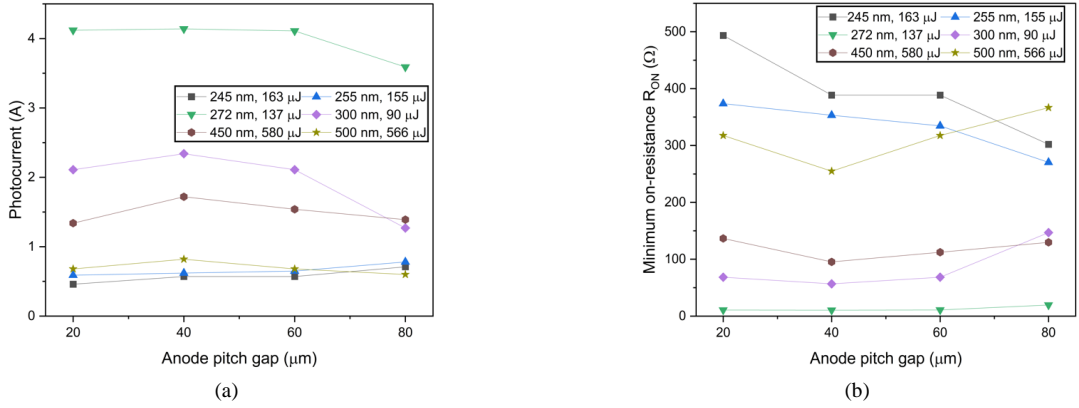

Fig. 4. (a) Peak photocurrent and (b) minimum on-resistance for different PCSSs when optically excited with sub- and above-bandgap wavelengths.

The transient photoresponses of Fe-doped β-Ga$_2$O$_3$ PCSS devices were analyzed by extracting the switch-on and switch-off times as functions of excitation wavelength. The switch-on time ($\tau_r$) is defined as the interval required for the photocurrent to reach its peak, while the switch-off time ($\tau_d$) corresponds to the decay from peak to zero current. As shown in Fig. 5(a), $\tau_r$ exhibits a distinct wavelength dependence, serving as a sensitive spectroscopic indicator of deep-level defect dynamics. Two regimes can be identified: an intrinsic regime ($\lambda < 270$ nm, $h\nu \gtrsim 4.6$ eV) characterized by direct band-to-band excitation (~1–1.5 ns), and an extrinsic regime ($\lambda > 270$ nm) dominated by slower (2.5–5 ns) trap-assisted transport. Specific spectral features in $\tau_r$ correspond to discrete defect levels: the delay peak near 280–300 nm arises from carrier excitation into Urbach tail states, while the feature around 350 nm is attributed to the ~3.5 eV optical transition associated with oxygen vacancies ($V_o$). The broader maximum near 450 nm likely reflects overlapping contributions from Fe$^{2+/3+}$ levels (~3.05 eV), gallium vacancies ($V_{Ga} \approx 3.0$ eV), and multi-step charge-transfer transitions involving Ir-related centers ($E_{Ir} > 2.0$ eV).

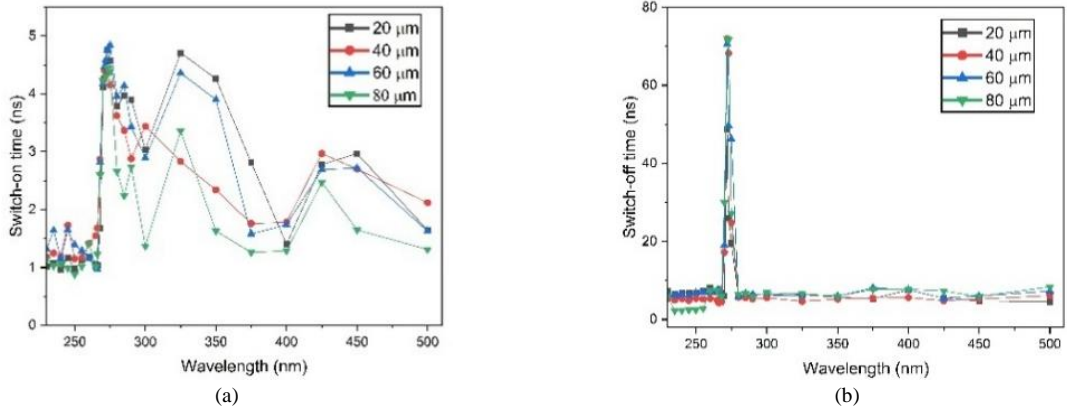

Fig. 5 (a) Switch-on and (b) switch-off times of the Fe-doped Ga$_2$O$_3$ PCSS with variation in excitation wavelength.

The decay dynamics, shown in Fig. 5(b), exhibit a complementary trend. Under high-energy excitation ($\lambda < 260$ nm), rapid current quenching occurs due to efficient electron–hole recombination following band-to-band absorption. For $\lambda > 300$ nm, excitation predominantly activates shallow defect levels, resulting in a weak and short-lived photocurrent. Notably, illumination within the 270–280 nm range photoionized a specific defect complex that forms long-lived hole traps, resulting in a persistent photocurrent and a markedly prolonged switch-off time. This persistent photocurrent is attributed to the photoionization of a defect complex likely associated with Fe-dopants or oxygen vacancies that functions as a long-lived hole trap [22-23].

In summary, excitation at 245 nm yields ultrafast rise times (< 2 ns) but relatively low responsivity (~5.5 × 10$^{-6}$ A-cm/W-kV). In contrast, excitation at 272 nm produces a relatively slower rise time (~4.5 ns) but yields a significantly higher responsivity of 3.89 × 10$^{-5}$ A-cm/W-kV, together with a low $R_{ON}$ of 10.4 Ω, indicating an efficient switching. The comparative 3D plots in Fig. 6 highlight the improved performance of the developed Fe-doped β-Ga$_2$O$_3$ PCSS relative to the previously reported UWBG photoconductive switches. As shown in Fig. 6(a), the device exhibits a markedly lower voltage-to-peak-current ratio, indicating efficient high-current conduction under moderate bias and sub-bandgap optical excitation. This enhanced current-driving capability—delivering ampere-level photocurrents with a R$_{ON}$ of ~10 Ω—contrasts sharply with previously reported Fe-doped β-Ga$_2$O$_3$ and diamond-based PCSS devices that required higher optical power or exhibited substantially higher $R_{ON}$.

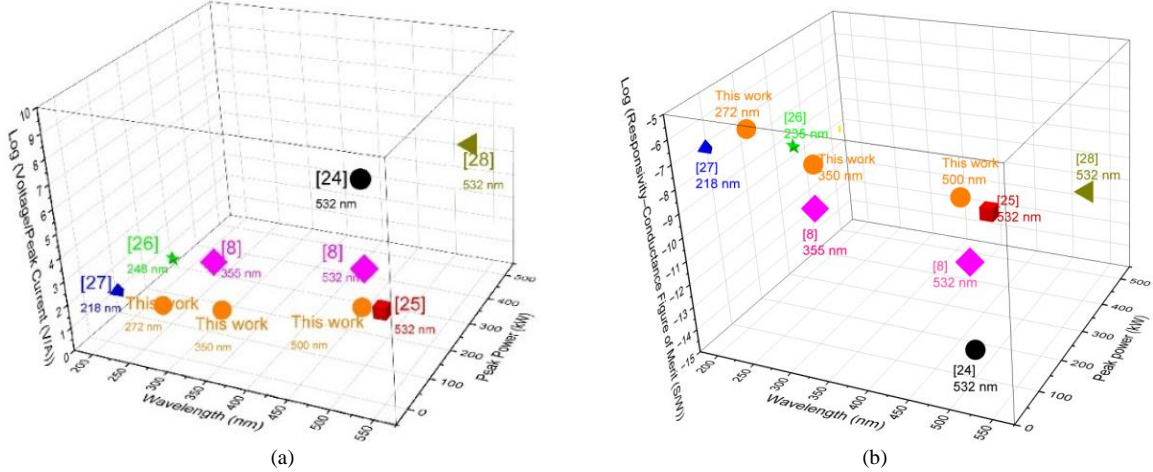

Fig. 6. Performance comparison of the Fe-doped β-Ga$_2$O$_3$ PCSS with previously reported UWBG photoconductive switches: (a) Voltage-to-peak-current ratio (where voltage refers to the power bias) and (b) responsivity–conductance figure of merit ($FoM_{RC}$) as a function of wavelength and peak power of the optical beam.

Complementarily, Fig. 6(b) presents a comparison of the responsivity–conductance figure of merit ($FoM_{RC}$) of previously reported Ga$_2$O$_3$- and diamond-based PCSS devices. A high $FoM_{RC}$ of 4.7 × 10$^{-6}$ S/W is demonstrated at 272 nm. Notably, this high $FoM_{RC}$ is achieved alongside a record-low on-resistance of 10.4 Ω, representing a new benchmark in the responsivity-conductance performance space. The improvement is attributed to an optimized electrode geometry, which provides a uniform electric field distribution and enhances carrier collection efficiency. Collectively, these results establish the present Fe-doped β-Ga$_2$O$_3$ PCSS as a new performance benchmark among UWBG photoconductive switches, combining high peak photocurrent, low on-resistance, and strong optical-to-electrical conversion efficiency within a compact and scalable device architecture. A logarithmic vertical scale is used to capture the wide dynamic range of responsivity and current, allowing both low- and high-level performance regimes to be clearly visualized within the same plots.

## 5. Conclusions

This work demonstrates the record responsivity–conductance figure of merit ($FoM_{RC}$) under sub-bandgap excitation in Fe-doped β-Ga$_2$O$_3$ PCSS. This performance enhancement is enabled through the concurrent optimization of the anode-grid pitch, serving as both the optical access window and the electric-field shaping element, and the excitation wavelength. The results show that excitation at 272 nm with a 40-μm contact pitch maximizes volumetric carrier generation and collection, yielding a peak photocurrent of 4.14 A, a record-low on-resistance of 10.4 Ω, and a corresponding $FoM_{RC}$ of 4.71 × 10$^{-6}$ S/W, thereby establishing a new benchmark for UWBG photoconductive switching technologies. For excitation at wavelengths longer and shorter than this optimum wavelength, device performance progressively diminishes due to increased optical transmission (weaker absorption) and surface-confined photocarrier generation (experiencing weak vertical field), respectively, both of which limit collection efficiency. Collectively, these results position Fe-doped β-Ga$_2$O$_3$ as a highly promising, cost-effective platform for next-generation high-power photoconductive switching, offering superior conductance and efficiency compared to recently reported UWBG-based PCSS counterparts.

**Acknowledgement**

The information, data, or work presented herein was funded in part by the Advanced Research Project Agency-Energy (ARPA-E), U.S. Department of Energy, under Award Number DE-AR0001879. Authors also acknowledge the support of the Nanotechnology Core Facility (NCF) at the University of Illinois Chicago.

## Author Declarations

Conflict of Interest: The authors declare no conflicts of interest.

## Data Availability

The data that supports the findings of this study are available from the corresponding author upon reasonable request.